\documentclass[11pt,preprint]{aastex}

\begin{document}

\title{Albedo heterogeneity on the surface of (1943) Anteros}

\author{Joseph Masiero\altaffilmark{1,2}}

\altaffiltext{1}{Institute for Astronomy, University of Hawaii, 2680 Woodlawn Dr, Honolulu, HI 96822}
\altaffiltext{2}{Jet Propulsion Lab, California Institute of Technology, 4800 Oak Grove Dr., MS 264-767,
  Pasadena, CA 91106, {\it Joseph.Masiero@jpl.nasa.gov}}

\begin{abstract}
  We have investigated the effect of rotation on the polarization of
  scattered light for the near-Earth asteroid (1943) Anteros using the
  Dual Beam Imaging Polarimeter on the University of Hawaii's $2.2~$m
  telescope.  Anteros is an L-type asteroid that has not been
  previously observed polarimetrically.  We find weak but significant
  variations in the polarization of Anteros as a function of rotation,
  indicating albedo changes across the surface.  Specifically, we find
  that Anteros has a background albedo of $p_v=0.18 \pm 0.02$ with a
  dark spot of $p_v < 0.09$ covering $<2\%$ of the surface. 
\end{abstract}

\section{Introduction}
As the last remnants from an epoch of accretive formation, asteroids provide
us windows into the composition and history of the inner Solar System.
Except for the few largest bodies, asteroids did not heat up enough via decay
of short-lived radionuclides or dissipation of gravitational potential energy
to undergo complete differentiation.  As such the minerals observed on
their surfaces capture the elemental and temperature history of the local
region of the protoplanetary disk at the time of their formation.  By
understanding asteroid surfaces we can directly probe those early disk
conditions.

As the illuminated cross section of an asteroid changes the observed
brightness fluctuates.  Given a large enough sample of data a full shape
model of a rotating body can be constructed \citep{mikko01} even though it
is unresolved.  Photometric surveys for asteroid light curves have set limits
on the composition and density of asteroids as a population
\citep{pravecAIII} and have estimated the average shape distribution of small
Main Belt asteroids \citep{talcs}.  All of these results however assume that
the light curve is dominated by the object's shape and that the entire
surface has a uniform composition and albedo.

It is possible to test for albedo variations using optical imaging
polarimetry, color variations, or even simply photometric variations
under the assumption of a regular shape \citep{akimov83}.  In the case
of polarimetry there are strong empirical correlations between the
albedo of an asteroid and both the slope of the polarization-phase
curve and the location of the minimum (negative) polarization
\citep[most recently:][]{cellino99}.  The polarization of light
scattered off of an atmosphereless body as a function of phase angle
depends on the distance between scattering elements and their index of
refraction \citep{muinonen89,muinonen02}.  Index of refraction is an
inherent mineralogical property and recent work has shown that the
inter-element scattering distance is likewise determined by the
surface chemistry \citep{masieroDust}.  It is not unexpected then that
asteroids of different spectral classification show different
polarization-phase curves \citep{muinonenAIII}, or that a
differentiated-then-broken object like (4) Vesta would show
polarization variations with rotation.

In almost every way investigated so far the asteroid Vesta stands out
as an interesting and unique object, and this is similarly the case
for albedo variation studies.  Although a handful of other asteroids
have weak detections of rotational modulation of their polarization
and thus albedos \citep[e.g.][etc.]{broglia92}, Vesta represents one
clear case of an object with polarization changes across its surface
caused by changes in composition
\citep{degewij79,lupishko88,broglia89}.  This, along with photometry,
spectroscopy and adaptive optic imaging, has lead to the current
interpretation of Vesta as a differentiated body that has undergone a
nearly-catastrophic impact event leaving a giant crater in its
southern hemisphere.  The crater reveals a now-solid mantle distinctly
different in color and composition from the original crust material
\citep{cellino87,thomas97}.

\citet{nakayama00} found rotational modulation of the polarization for
the asteroid (9) Metis with amplitude similar to what is observed for
Vesta.  Metis is a $D \sim 180~$km asteroid that may have two large
spots of significantly higher albedo ($p_v\sim0.24-0.28$) than the
background material ($p_v=0.14$).  The authors find that both bright
areas are on the leading (for prograde rotation) or trailing (for
retrograde rotation) faces of the model ellipsoid \citep{mitchell95}.
The cause of albedo heterogeneity across the surface of objects
smaller than Vesta is still undetermined.  It is possible that
non-disruptive collisions with impactors of different composition can
leave localized deviations from the average mineralogy.
Alternatively, a late formation with a history free of melting may
preserve the varied composition of the protoplanetary disk.  However
this theory is complicated by recent work showing that asteroids
likely were born big, and most objects smaller than a few hundred
kilometers in diameter should be collisionally-created fragments
\citep{morbi09}.

Identifying albedo variations for small asteroids allows us to evaluate the
accuracy of the assumption that flux changes are solely dependent on shape.
This has important implications for results based on this assumption,
especially shape models.  Additionally we can also quantify the effect of
collisions between small bodies in determining an asteroid's local regolith
properties.

All asteroid polarization-phase curves follow the same general trend
with increasing phase angle: zero polarization at zero phase, becoming
negative to some minimum value and then increasing in an approximately
linear fashion.  Note that as is standard for Solar system polarimetry
the reference direction for the angle of polarization is aligned with
the vector perpendicular to the plane of scattering such that
``positive'' and ``negative'' polarization are defined as
perpendicular and parallel to the scattering plane, respectively.  The
results presented here follow this convention.  Each
polarization-phase curve displays three distinguishing values used to
classify its properties: the minimum negative polarization
($P_{min}$), the phase angle at which the polarization returns to zero
(the inversion angle, $\alpha_0$) and the linear slope of the
polarization-phase relation beyond the inversion angle ($h$).  Making
use of the albedo-polarization relation from \citet{cellino99},
\begin{eqnarray}
\label{eq.slope8}
\log p_v & = & (-1.12\pm0.07) \log h - (1.78\pm0.06)
\end{eqnarray}
(where $p_v$ is the geometric V-band albedo) we can use imaging polarimetry
to test for changes in polarization that directly indicate albedo
heterogeneity across an asteroid's surface.

\section{Observations and Discussion}

Changes in the polarization of the scattered light across the surface of an
asteroid will be small even in the best-case scenarios.  To obtain a
significant measurement of the largest of these variations we require an
instrument that can attain better than $0.1\%$ polarization accuracy.  Our
study made use of the Dual Beam Imaging Polarimeter (DBIP) located on the
University of Hawaii's $2.2~$m telescope on Mauna Kea, Hawaii \citep{dbip}.
DBIP uses a double-calcite Savart plate in series with a quarter-wave and a
half-wave retarder to simultaneously measure linear and circular polarization
with accuracy better than $0.1\%$ \citep{dbip2}.  DBIP uses a $g'+r'$
filter with a bandpass of $400-700~$nm.  While asteroid polarization
does depend on color \citep{cellino05} changes are usually small in
this wavelength range and typically within measurement errors.

Observation of our target asteroid were supplemented with polarized
and unpolarized standards to confirm consistency of setup, stability
of the instrument, and accuracy of the measurements.  Standards were
taken from \citet{fossati07} as well as the standard list for
Keck/LRISp\footnote{http://www2.keck.hawaii.edu/inst/lris/polarimeter/polarimeter.html}
which includes the {\it Hubble Space Telescope} polarimetric standards
\citep{hubbleSTD}.  These measurements all verified that the errors
were within the range expected from previous calibrations.

As albedo is related to the polarization-phase slope $h$, for a given
albedo variation the respective polarization change will be larger
when observed at higher phase angles (for $\alpha > \alpha_0$).
Geometric restrictions prevent Main Belt asteroids (MBAs) from ever
reaching phases angles larger than $\alpha\sim30^\circ$, but
near-Earth asteroids (NEAs) pass closer to Earth and so can reach much
larger phase angles.  For this reason, NEAs are preferred targets when
looking for albedo variations.  At high phases the polarization of
scattered light takes on a linear trend that increases up to the level
of $\sim5-10\%$ polarized depending on surface mineralogy.  These
large polarizations mean that any variation with rotation at high
phases can be easily interpreted as changes in the integrated surface
albedo using the slope-albedo relation (Eq~\ref{eq.slope8}) and that
the absolute value for the range of the albedo can be determined.

From July to September of 2009 the NEA (1943) Anteros passed through
phase angles of $16-40^\circ$ all while brighter than $V=17~$mag
presenting a prime opportunity to measure the polarization, slope and
albedo with high accuracy.  The optical/NIR spectrum of Anteros
displays a spectral slope comparable to S-types but with a muted
$1~\mu$m absorption band resulting in a classification of L-type
\citep{binzel04}.  With a measured period of $P=2.8695\pm0.0002~$hr
and a single-peaked photometric light curve with amplitude
$A=0.09~$mag \citep{pravec98}, Anteros is an excellent target to test
for rotational variation in polarization and albedo in a few nights of
observing.  In particular, a single-peaked low-amplitude light curve
indicates a shape very close to spherical.  (\citet{pravec98} found an
amplitude of $0.09~$mag across phase angles ranging from
$19^\circ<\alpha<32^\circ$, resulting in a shape approximation of
$a/b<1.1$).

\begin{figure}[ht]
\begin{center}
\includegraphics[angle=-90,scale=0.5]{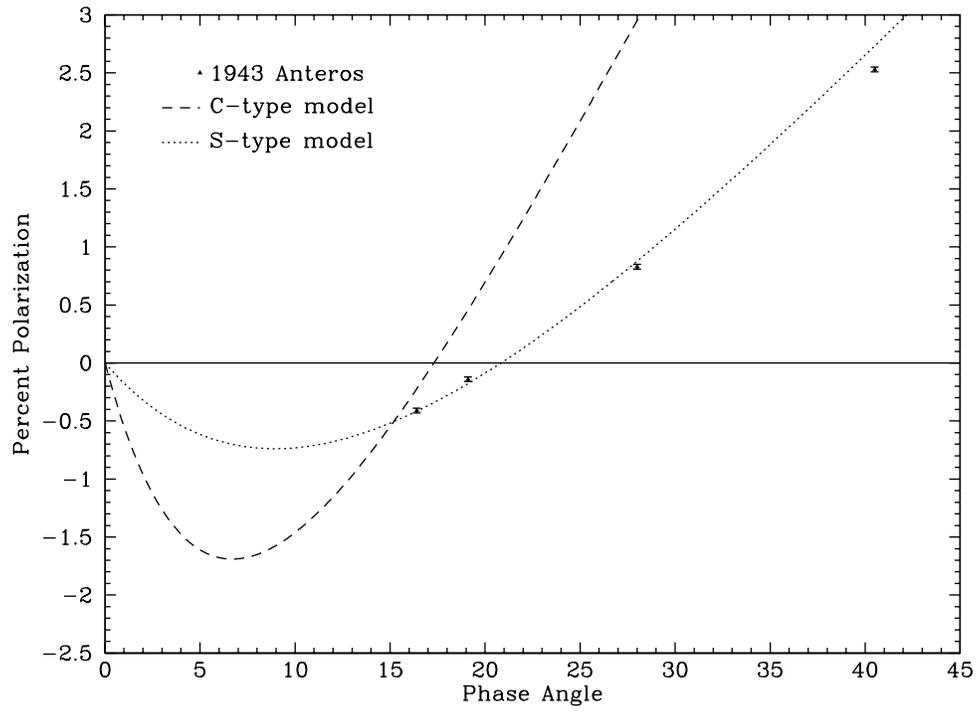}
\protect\caption[Polarimetry of (1943) Anteros compared to S-type and C-type model curves]{
New observations of the polarization of (1943) Anteros along with
model curves for S-type (dotted) and C-type (dashed), for reference.
Note the errors on the percent polarization are comparable to the size
of the points.
}
\label{fig.polphaseA}
\end{center}
\end{figure}

In Table~\ref{tab.anteros} we present our polarimetric observations of
Anteros.  Included for each night are the V magnitude, exposure time,
number of 6-exposure polarimetry measurements acquired ($n_{meas}$),
phase angle, ecliptic longitude, summed linear polarization of all
measurements and linear polarization angle with respect to the vector
perpendicular to the scattering plane ($\theta_p$).  No significant
circular polarization was detected on any of the nights.  The average
nightly polarizations are shown in Fig~\ref{fig.polphaseA} along with
generic model polarization-phase curves for typical S-type asteroids
(dotted) and C-type asteroids (dashed).  The model curves were made
using the linear-exponential modeling technique presented by
\citet{muinonenLEmod} and fitted by-eye to the data shown in Fig 1 of
\citet{muinonenAIII}, to act as useful approximations.  The constants
used in this case were, for the C-type model: $P_a = 5.5$, $P_d = 6$,
$P_k=0.3$ and for the S-type model: $P_a = 4.3$, $P_d = 12$,
$P_k=0.17$.

The polarization of Anteros is clearly most closely related to an
S-type polarization curve as expected from its spectral features.  We
measure for Anteros an inversion angle of $\alpha_0 = 20.3 \pm
0.3^\circ$ and a slope beyond the inversion angle of $h = 0.122 \pm
0.001$.  Both $\alpha_0$ and $h$ (and their respective errors) were
found by conducting a least-squares minimization fit of a line to the
four nights of data.  As the data span a large range of phases and
have small individual errors, the resultant error on $h$ is small.
Following Eq~\ref{eq.slope8} we derive a bulk albedo of $p_v = 0.175
\pm 0.002 \pm 0.02$ (errors relative and absolute, respectively).
Note that the limiting error on albedo ($\sigma_{abs} = 0.02$) derives
from the uncertainty on the constants in Eq~\ref{eq.slope8} and will
affect absolute albedo measurements.  This does not apply to relative
comparisons between measured albedos, which have an error of
$\sigma_{rel} = 0.002$ in the above case.  For all observations
reported here, the calibration error on Eq~\ref{eq.slope8} is greater
than the noise error by nearly an order of magnitude, and thus
dominates the final error on the measured albedos.  (All errors
reported in this paper are $1~\sigma$.)
 
Following \citep{muinonen02} polarization can be approximated as
\[P \sim \frac{\alpha^2}{2 n} - \left(\frac{n-1}{n+1}\right)^2 \frac{(k~d~\alpha)^
2}{2~[1+(k~d~\alpha)^2]} \]
and solving this for the case of zero polarization at the inversion
angle we find
\begin{eqnarray}
\label{eq.index}
\alpha_0 \sim \sqrt{n \left(\frac{n-1}{n+1}\right)^2 -\frac{1}{(kd)^2}}
\end{eqnarray}
where $k = 2\pi/\lambda$, $d$ is the inter-element scattering
distance, and $n$ the index of refraction.  For a central wavelength
of $0.55~\mu$m for DBIP and a typical scattering distance for NEAs of
$d \sim 4~\mu$m \citep{masieroDust} the second term on the right in
Eq~\ref{eq.index} is negligible, and the index of refraction can be
determined for a given inversion angle.  For Anteros we find an index
of refraction of $n \sim 1.74$.

\begin{figure}[ht]
\begin{center}
\includegraphics[scale=0.5]{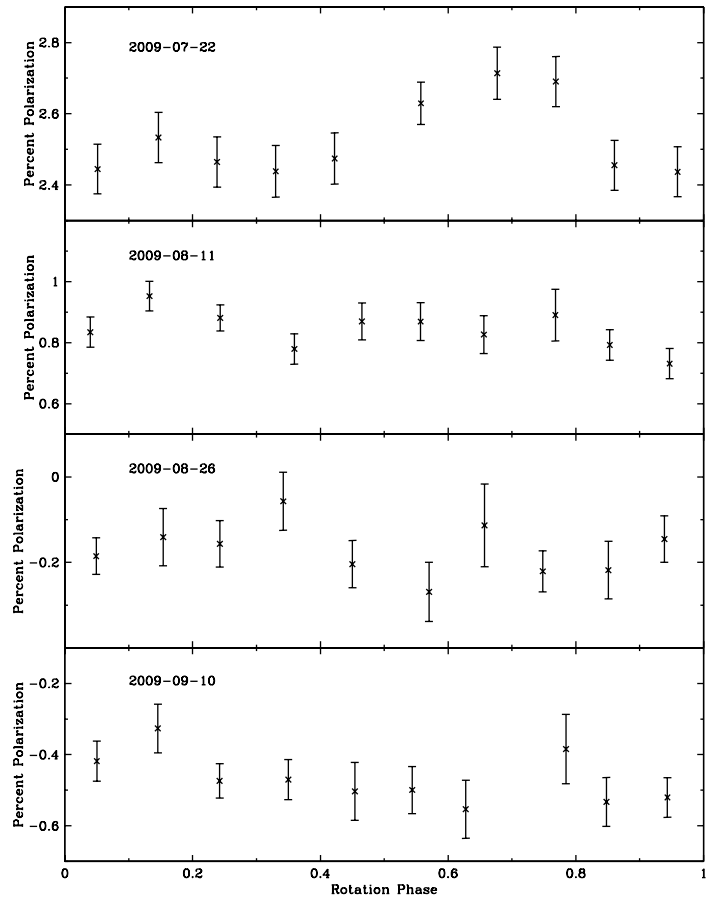}
\protect\caption[Polarization variation of (1943) Anteros as a function of rotation phase for 4 nights]{
Polarization of Anteros as a function of rotation phase.  Observations within $0.1~$phase-wide bins have been co-added to reduce errors.  UT date of each observation set is listed.  
}
\label{fig.polrotA}
\end{center}
\end{figure}

In Fig~\ref{fig.polrotA} we show the nightly polarization light curves for
Anteros.  Observations have been wrapped to the measured $2.8695~$hr
photometric period and all measurements within a $0.1~$phase bin have been
co-added to reduce measurement error.  The zero point for rotation phase was
chosen arbitrarily on the first night.  When all nights of data are wrapped
to this zero-point the error on the period translates to a phase error of
$\pm 0.01~$rotations between each observing night.  Thus, features at
specific phases can be compared across nights.

We find weak variation in the polarization at a $4\sigma$ significance
level for the night of 2009-07-22 with an amplitude of $0.3\%$ when
comparing the data in the rotation phase range of $0.6-0.8$ to the
data between rotation phases $0.8-0.4$.  These polarization changes
most likely indicate a variation in the albedo across the surface of
Anteros.  The amplitude of the polarization variation scales with the
absolute polarization so it is not surprising that the other observing
nights at lower phases show no clear variation, e.g. a variation with
amplitude of $P=0.3\%$ at a phase of $\alpha=40.5^\circ$ would be
expected to have an amplitude of $P=0.1\%$ at a phase of
$\alpha=28.0^\circ$ which is below our threshold for significant
detection.  Additionally, changes in the observing geometry between
observations could cause an area that was observable on the night of
2009-07-22 to have a reduced visibility on subsequent observations, or
even be beyond the horizon.  However even in the most extreme case,
where the rotation axis is parallel the to plane of Earth's orbit, the
line-of-sight vector only moves a total of $\sim18^\circ$ with respect
to the rotation axis over the dates of the observations (this is
equivalent to the change in ecliptic longitude).  Though this could
account for the changes in polarization under specific circumstances,
it is not the most likely scenario.

It has been suggested that the change in polarization alternatively
could be due to an extreme topographical feature that deviates
significantly from the surrounding area.  At the high phases at which
we observed Anteros first-order scattering is dominant and so we may
simply apply basic scattering properties to the surface (e.g. the
angle of incidence and angle of scattering are equal, etc).  Thus the
light we observe necessarily must have been scattered by planes on the
surface normal to the vector that bisects the phase angle.  Even on
unusual surfaces, microroughness will provide the appropriate
scattering facets.  In the extreme argument, should the plane be a
perfectly flat surface (e.g. a mirror) it will scatter no light to the
observer when away from a perfect alignment.  This will decrease the
overall flux but not change the percent of the received flux polarized
by the surrounding area.  If aligned perfectly, the plane will still
behave polarimetrically as the underlying material it is made of,
polarizing the same fraction of the light as determined by its albedo.
Thus even in case of extreme topography, percent polarization
measurements will only be sensitive to the underlying material
composition.

Changes in the absolute level of polarization as the phase angle
changes nightly precludes wrapping all four nights of observing onto a
single rotation phase.  However we can correlate features at similar
rotation phases across nights and from this fit different polarization
slopes to different locations on the asteroid.  We use the phase range
of $0.6-0.8$ to represent the peak polarization and the phase range of
$0.8-0.4$ (wrapped) to represent the baseline background polarization
(as determined from the 2009-07-22 observing night).  We interpret the
phase range of $0.4-0.6$ as a transition region (see below) and do not
include it in either measurement.  Using the summed maximum and
baseline values for the first night as location benchmarks we
determine the absolute change in albedo across the surface of Anteros.

We find a background surface albedo of $p_v = 0.181 \pm 0.002 \pm
0.02$ with a single spot of much lower albedo.  Note that this value
does not vary significantly from the albedo of $p_v = 0.13 \pm 0.03$
found from radiometric modeling \citep{veeder81} or the one published
in the compilation by \citet{chapman94} of $0.17$ (no error given).
We measure an upper limit to the albedo for the dark area of $p_v =
0.160 \pm 0.004 \pm 0.02$ however this feature is unresolved and thus
the albedo measurement assumes coverage of the maximal possible area
allowable by the observing geometry ($44\%$ of the total surface area
for 2009-07-22, corresponding to a projected area of $3.7~$km$^2$).
It is likely that the dark spot covers a much smaller area with a much
lower reflectance.

Reflected polarization from a mottled surface will mix when unresolved to
give a value between the two extremes.  For the case of Anteros on the night
of 2009-07-22, taking a background polarization for $\alpha=40.5^\circ$ of
$P_{bkg} = 2.46\%$ (the mean of the baseline polarization value for the night) and a
peak measured polarization of $2.70\%$ we find that the true percent
polarization of the dark spot ($P_{dark}$) on that night can be described as
\[P_{dark} * C + P_{bkg} (1-C) = 2.7\] 
where $C$ is the fraction of the projected illuminated surface that
the dark region covers.  This is simply because the polarized light
from the dark area is diluted by the signal from any background
material also visible.  This relation can be simplified to:
\begin{eqnarray}
\label{eq.coverage}
P_{dark} = \frac{0.24}{C} + 2.46
\end{eqnarray}

The 2009-07-22 data plotted in Fig~\ref{fig.polrotA} show a gradual build up
in the polarization value with a rapid falloff from the peak level to the
base level over slightly more than one tenth of a rotation.  This is likely
due to the dark region rising over the horizon as seen from Earth and then
passing from the lit side quickly across the terminator and thus out of
illumination.  This scenario would require Anteros to have a prograde
rotation state.  From the rapid falloff we can calculate that the maximum
size of the dark feature along the direction of rotation is $<0.7~$km
assuming it is located on the equator of the asteroid.  If the spot is
not on the equator this argument would derive a smaller value for the size.

Using this size as the diameter of a circular crater this corresponds
to a total projected surface coverage of the dark spot of $C<11\%$
(equivalent to $<2\%$ of the total unprojected surface area, assuming
the rotation pole is parallel to the plane of the sky), giving a
polarization value for this area of $P_{dark}=4.7\%$.  If instead the
rotation axis is inclined to the plane of the sky, the projected area
would be smaller.  This would increase the required value of
$P_{dark}$ which would cause the calculated value for the albedo (see
below) to decrease.  Thus the assumptions made here represent the
'brightest-case' scenario for the spot.

\begin{figure}[ht]
\begin{center}
\includegraphics[angle=-90,width=\textwidth]{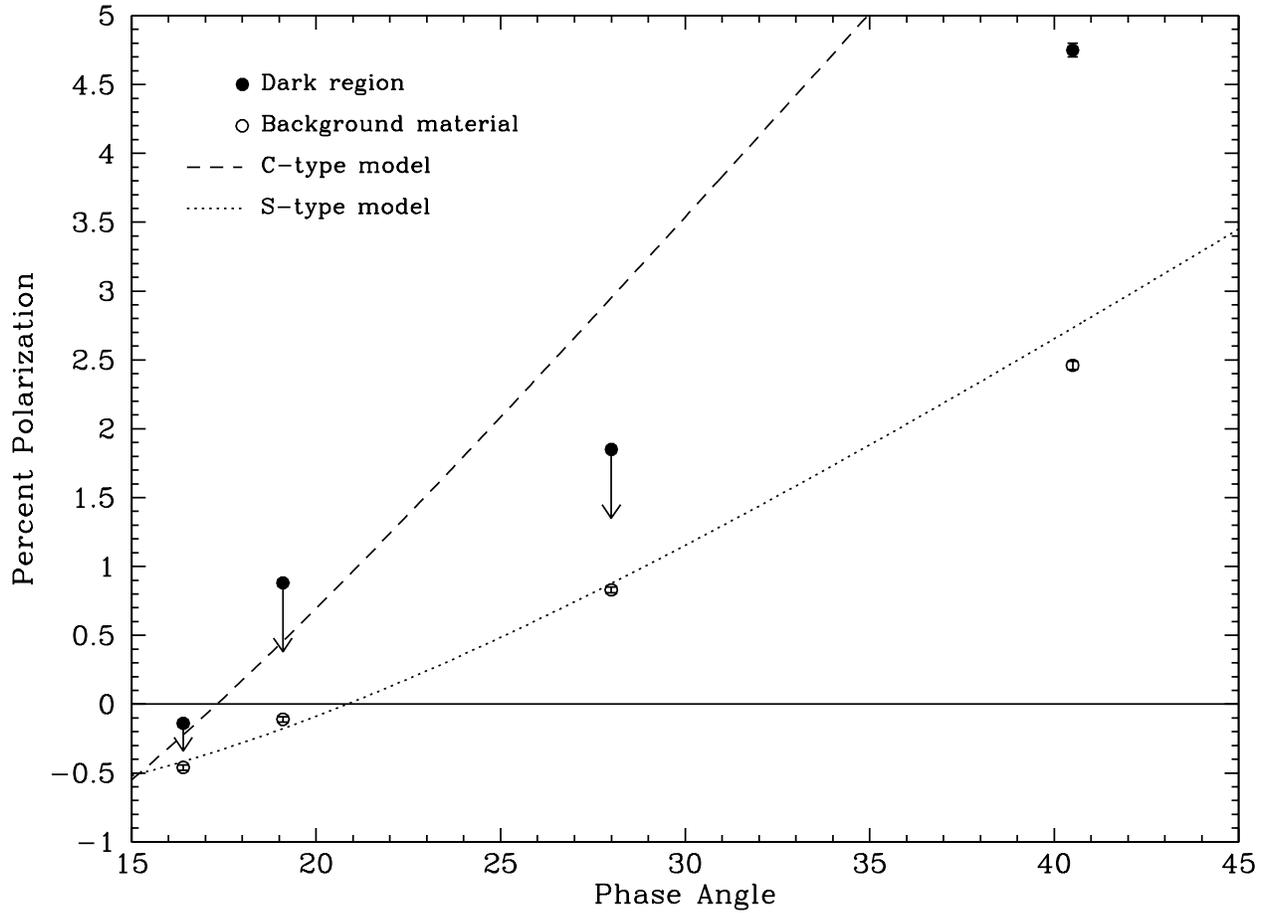}
\protect\caption[Polarization of the background and the dark region of (1943) Anteros]{
Polarization of the background and the dark region of (1943) Anteros.  The values of the polarization for the dark region for phase angles below $40.5^\circ$ are upper limits.  Note that the error bars on the points are comparable to their size.
}
\label{fig.lightdark}
\end{center}
\end{figure}

As the polarization for the other three nights is consistent at all
phases we can only set an upper limit for the value of $P_{dark}$ on
those nights based on a maximum unresolved change in percent
polarization of $0.1\%$.  Using the analouges of Eq~\ref{eq.coverage}
for the other nights we find limits of $P_{dark} < 1.85\%$ for the
night of 2009-08-11, $P_{dark} < 0.88\%$ for 2009-08-26, and $P_{dark}
< -0.14\%$ for 2009-09-10.  We show the polarization for the dark
region separated out from the background material in
Fig~\ref{fig.lightdark} as well as the same S- and C-type generic
models from Fig~\ref{fig.polphaseA}.  Using these values we calculate
a lower limit on the slope and thus an upper limit on albedo.  The
albedo of the dark spot is limited to $p_v<0.09$.

\section{Conclusions}

Using DBIP on the University of Hawaii's $2.2$~m telescope we have
investigated the NEA (1943) Anteros for surface heterogeneity.  We
find that Anteros shows significant polarimetric variation as a
function of rotation at high phase angles, implying an albedo gradient
and corresponding surface composition variations.  We determine that
Anteros has a base albedo of $p_v = 0.181 \pm 0.002 \pm 0.02$ (errors
are relative and absolute, respectively) consistent with literature
values as well as a dark spot of albedo $p_v < 0.09$ covering
$<2\%$ of its surface. 

A single small asteroid showing albedo variations does not invalidate
the assumption that shape dominates the light curves of these bodies.
Indeed it is clear from spacecraft visits that shape does play an
important role in determining the reflected flux.  Additionally,
albedo variations for most asteroids appear to be small and localized.
However the potential for albedo variations for small asteroids can
not be discounted outright.  

It is currently unclear what processes could cause localized albedo
changes across the surfaces of small asteroids.  Further polarimetric
studies of small NEAs are thus necessary to constrain the frequency of
albedo and composition changes across the surfaces of these bodies.
Once an account the population is established evolutionary pathways to
the creation of these features can be explored.

\section*{Acknowledgments}

J.M. was supported under NASA PAST grant NNG06GI46G.  The author would
like to thank Rob Jedicke and Alan Tokunaga for providing comments on
the manuscript, as well as V. Rosenbush and an anonymous referee for
helpful reviews that improved the paper.  The author wishes to
recognize and acknowledge the very significant cultural role and
reverence that the summit on Mauna Kea has always had within the
indigenous Hawaiian community. I am most fortunate to have the
opportunity to conduct observations from this sacred mountain.

%
%

\newpage

\begin{table}[ht]
\begin{center}
\caption{Polarimetry of (1943) Anteros}
\vspace{1ex}
\noindent
\begin{tabular}{cccccccc}
\tableline
UT Obs Date  & V mag & $t_{exp}$ (sec) & $n_{meas}$ & $\alpha$  &
Ecliptic Long & Lin $\%~$Pol & $\theta_p$ \\
\tableline
2009-07-22 & 16.6 & 130 & 21 & $40.5^\circ$ &$356.9^\circ$ & $2.53 \pm 0.02$ & $178.8 \pm 0.3^\circ$ \\
2009-08-11 & 16.3 & 130 & 26 & $28.0^\circ$ & $352.9^\circ$&  $0.83 \pm 0.02$ & $0.8 \pm 0.6^\circ$ \\
2009-08-26 & 16.1 & 90 & 27 & $19.1^\circ$ & $346.3^\circ$& $-0.14 \pm 0.02$ & $86 \pm 4^\circ$   \\
2009-09-10 & 16.3 & 100 & 30 & $16.4^\circ$ & $339.3^\circ$& $-0.41 \pm 0.02$ & $94 \pm 1^\circ$  \\
\hline
\end{tabular}
\label{tab.anteros}
\end{center}
\end{table}

\clearpage

\end{document}